# Metal artefact reduction sequences for a piezoelectric bone conduction implant using a realistic head phantom in MRI

*G. Fierens, PhD[1,2,3*], J. Walraevens, PhD[2], R. Peeters, PhD[4], C. Glorieux, PhD[1,] N. Verhaert, MD, PhD[3,5]*

[1] Laboratory of Soft Matter and Biophysics, Department of Physics and Astronomy, KU Leuven, Celestijnenlaan 200D, B-3001 Heverlee, Belgium,

[2] Cochlear Technology Centre, Schaliënhoevedreef 20I, B-2800 Mechelen, Belgium,

[3] Research group Experimental Otorhinolaryngology, Department of Neurosciences, KU Leuven, Herestraat 49, B-3000 Leuven, Belgium.

[4] University Hospitals Leuven, Department of Radiology, Herestraat 49, B-3000 Leuven, Belgium

[5] University Hospitals Leuven, Department of Otolaryngology, Herestraat 49, B-3000 Leuven, Belgium

*Corresponding author: Guy Fierens, Laboratory for Soft Matter and Biophysics, Department for Physics and Astronomy, Celestijnenlaan 200D, Box 2416, 3000 Leuven. Email: guy.fierens@kuleuven.be

Grant support: This work has been partly supported by Flanders Innovation and Entrepreneurship [HBC.2018.0184] and Cochlear Limited. NV has a senior clinical investigator fund [FWO 1804816N].

# Abstract

Industry standards require medical device manufacturers to perform implant-induced artefact testing in phantoms at a pre-clinical stage to define the extent of artefacts that can be expected during MRI. Once a device is commercially available, studies on volunteers, cadavers or patients are performed to investigate implant-induced artefacts and artefact reduction methods more in-depth.

This study describes the design and evaluation of a realistic head phantom for pre-clinical implant-induced artefact testing in a relevant environment. A case study is performed where a state-of-the-art piezoelectric bone conduction implant is used in the 1.5T and 3T MRI environments. Images were acquired using clinical and novel metal artefact reducing (MARS) sequences at both field strengths. Artefact width and length were measured in a healthy volunteer and compared with artefact sizes obtained in the phantom.

Artefact sizes are reported that are similar in shape between the phantom and a volunteer, yet with dimensions differing up to 20% between both. When the implant magnet is removed, the artefact size can be reduced below a diameter of 5 cm, whilst the presence of an implant magnet and splint creates higher artefacts up to 20 cm in diameter. Pulse sequences have been altered to reduce the scan time up to 7 minutes, while preserving the image quality.

These results show that the anthropomorphic phantom can be used at a preclinical stage to provide clinically relevant images, illustrating the impact of the artefact on important brain structures.

# Introduction

Research into image artefacts in MRI has been partly focused on reducing image artefacts induced by implantable medical devices. Manufacturers of those devices comply to industry standards, which require them to perform preclinical experiments in order to show the worst-case image artefacts in the device guidelines [1]–[3]. Research in the past years has focused on reducing implant-induced artefacts, with the goal to increase the diagnostic value of MRI in patients with different types of active or passive implants. This has been tackled by optimizing metal artefact reduction sequences (MARS) [4]–[8].

A relatively new implantable technology can be found in active transcutaneous bone conduction implants (ATBCIs), which have been introduced on the market in 2012. These ATBCI are an implantable hearing solution that partly restores the hearing function of patients suffering from conductive or mixed hearing loss [9]. They bring the advantage of being mostly implantable, keeping the covering skin layers intact. This in contrast to passive percutaneous devices, which due to their percutaneous nature created the risk of skin infection for the patient [10]. Most hearing implants contain a permanent magnet to align an internal and external coil, which is required to couple the inner part with the externally worn speech processor. In addition to the magnet, most ATBCIs contain electromagnetic actuators, such as the Bonebridge [11] or the "bone conduction implant" [10], [12].

More recent ATBCI are based on piezo-electrical actuator technology [9], [13]. This technology also comes with different behaviour with respect to mutual interactions in the magnetic resonance imaging (MRI) environment compared to the electromagnetic variants. As the use of MRI has been increasing in the past decades, implant manufacturers have been working towards improving MRI guidelines for their devices in order to allow for any post-operative examinations[3].

Research in artefact reduction for these ATBCI feature a variety of phantoms, cadavers or in-vivo data. It is clear that in-vivo data present the most clinically relevant situation, yet clinical data can only be gathered when a device has received regulatory approval with regard to ethical considerations. Studies using clinical data nonetheless provide valuable insights concerning artefact reduction[5] or artefact localization and/or size[14]–[18]. For pre-clinical studies, researchers have to rely on either cadavers or phantoms. Post-mortem degradation of a specimen can introduce technical challenges, although Thiel embalming of cadavers provides a stable quality specimen as indicated by the research performed on the Med-El BoneBridge[11], [19], [20]. Fresh cadavers have however also been used when investigating the effect of the presence of an implant magnet on image artefacts [8], [21]–[23].

The challenges involved in *in-vivo* or *post-mortem* research in this field have pushed researchers to develop phantoms for pre-clinical studies. The industry standard for image artefact testing, ASTM F2119-07 (2013)[1] refers to the use of a water bath containing a reference grid, whilst other traditional phantoms consist of cylindrical containers filled with tissue-mimicking gels or liquids[6], [24]. The design and use of anthropomorphic MRI phantoms has been mostly focused on mimicking relaxation times to allow validating image processing strategies, however without a clear focus on the spatial accuracy in the brain region that would be relevant for brain imaging in the presence of implants. A recent overview of the available models is provided in the review paper by Crasto et al. [24]. In the work of Shmueli et al., a phantom was built for imaging on a scanner with a field strength of 4.7T, which contained some air-filled cavities and bone structures [25] whilst Altermatt et al. described a phantom design where both the grey and white matter were present based on the segmentation of an MRI dataset[26]. The degree of sharpness of the brain morphology present in this phantom was however limited, in part due to the fact that the brain was moulded and that no central structures like the ventricles were present. Wood et al. on the other hand presented an anthropomorphic head phantom, where the brain was constructed of a combined grey and white matter structure and a separate brainstem structure [27].

The aim of this work is two-fold. The first aim is to develop an anthropomorphic phantom featuring bone and soft tissue mimicking structures, including a brain-like structure that shows grey matter, white matter and cerebrospinal fluid in the ventricles. In order to assess the representativeness of the design in pre-clinical testing, the phantom is compared to an *in-vivo* test case. In a second research aim, the phantom is then used to investigate the use of metal artefact reduction sequences in a case study for the Cochlear™ Osia® OSI200 system, a novel ATBCI that recently has received regulatory approval.

## Materials and methods

### Phantom design

The main structure of the head phantom was based on an artificial skull with a mandible and C1-C6 vertebrae (model 1345-32, Sawbones Europe AB, Limhamm, SE). The anteroposterior length of the skull is 17.6 cm, corresponding to a male skull[28]. Soft-tissue structures consisted of the eyes and brain. Eyes were fabricated using hollow spheres with a diameter of 30 mm, which were filled with silicone to provide a realistic contrast during MRI. Spheres were 3D printed on a stereolithography (SLA) 3D printer using standard resin (Form 2, Formlabs, Somerville, Massachusetts, USA).

A model for the brain consisting of grey and white matter was segmented out of an open MRI dataset [29] using an open-source software package DeVIDE [30]. Both tissue types were segmented out of the MRI dataset using a custom algorithm based on thresholding and mesh smoothing. Segmented models were then exported as STL files for further editing in Meshmixer (Autodesk, Mill Valley, California, USA). Editing included repairing errors in the STL models as well as smoothing and resizing to ensure a proper fit in the intracranial volume of the available skull model. After resizing, both models were hollowed-out until a wall thickness of 1 mm was achieved. Drainage holes were added in both models to allow filling up the model with a contrast agent to create a realistic contrast on MRI images before being 3D printed.

In order to determine the proper contrast agent concentration, an experiment was performed where 6 containers of different solutions were prepared with either $MnCl_2$ (CAS: 13446-34-9) or $CuSO_4$ (CAS: 7758-98-7) as a contrast agent. The relaxivity as a function of concentration was then defined for T1 and T2 relaxation based on inversion recovery turbo spin-echo (2D IR-TSE, horizontal slice, 0.6 x 0.6 mm² resolution, slice thickness = 6 mm, TR = 4s, TE = 15 ms, TI = 50/100/200/300/400/600/800/1000/1500/2000/3000 ms, Echo train length = 5) and multi-echo mapping (2D TSE, horizontal slice, 1.1 x 1.1 mm² resolution, slice thickness = 6 mm, TR = 4s, TE = 10 to 320 ms in 10 ms increments, Echo train length = 32) sequences in the 1.5 T environment, respectively. Relaxivity curves were obtained using the 2-parameter algorithm outlined in the work of Thangavel and Saritas [31]. A concentration of 0.025 mM $MnCl_2$ was chosen to match the relaxivity of the grey matter, a concentration of 0.044 mM $MnCl_2$ was used to match the relaxivity of white matter and a concentration of 1.1 mM $CuSO_4$ was used to match the relaxivity of cerebrospinal fluid according to Haacke et al. [32]. The brain model was filled with the $MnCl_2$ solution whilst a water bath was filled with the $CuSO_4$ solution. Figure 1 below depicts the head phantom containing the $MnCl_2$-filled brain placed into a water bath before the $CuSO_4$ contrast agent was added.

The algorithm used to segment the brain, the obtained relaxivity curves, as well as the 3D models for the soft tissue structures, are presented in the Supplementary Materials of this work.

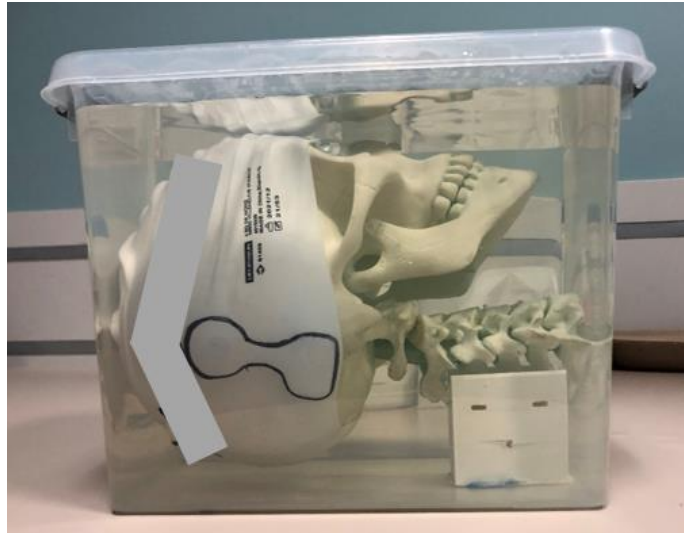

*Figure 1: Head phantom containing the MnCl2-filled brain placed in a water bath before the CuSO4 contrast agent was added. The brand name of the swimming cap is blurred out.*

## Study design

A BI300 bone screw (Cochlear Ltd., Sydney, AU) was implanted in the right temporal bone of the phantom at a distance of 4 cm posterior to the centre of the ear canal according to the manufacturer's surgical guidelines. The Cochlear™ Osia® OSI200 system was then mounted onto the bone screw with the inductive coil pointing upward as depicted in Figure 2a.

In order to compare images acquired with the phantom with images from a clinical case, an implant was also attached to the right side of the head of a healthy volunteer (male, 29 years; Figure 2b). A written informed consent was signed by the volunteer before the start of the study. Both for the phantom and the volunteer the device was additionally kept in place by placing it under a swimming cap as indicated in Figure 2. The use of a swimming cap was inspired by Dewey et al. [33]. For 1.5T MRI when the magnet was kept in place, the Cochlear™ Osia® MRI kit was also used in line with the manufacturer guidelines.

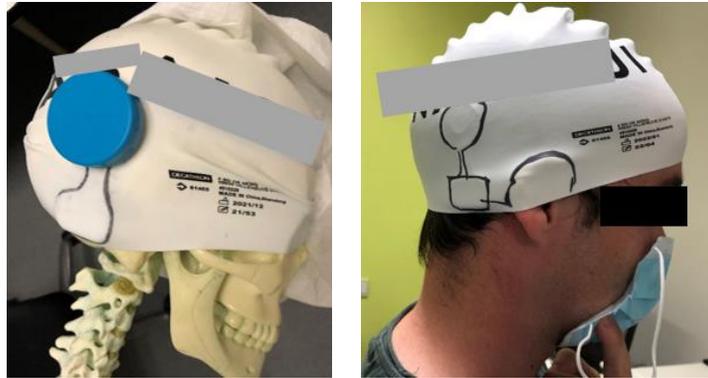

*Figure 2: Position of the implant on the phantom (a) and the head of a healthy volunteer (b) below a swimming cap to keep the implant in place. The position of the underlying implant is depicted using a permanent marker on the swim cap. The brand name of the swimming cap is blurred out.*

## Image acquisition

Images were acquired on clinical cylindrical bore MRI scanners with field strengths of 1.5 T (Intera 1.5T, Philips Healthcare, Best, NL) and 3 T (Achieva 3T dStream, Philips Healthcare, Best, NL). Datasets were first acquired in the 3T environment where the implant magnet was replaced with a non-magnetic plug. Next, images were acquired using the non-magnetic plug in the 1.5T environment, before replacing the non-magnetic plug with a magnet. The Cochlear MRI kit was then applied in accordance with the manufacturer's MRI guidelines. An overview of the described test conditions is provided in Table 1.

*Table 1: Implant configurations in both MRI scanners*

| MRI Field strength | Implant magnet | MRI Kit |
|---|---|---|
| 1.5 T | Yes | Yes |
|  | No | No |
| 3.0 T | No | No |

Pulse sequences were defined within the constraints posed in the device guidelines, including manufacturer-recommended pulse sequences and a number of examinations in order to represent a collection of the most clinically relevant examinations used in the hospital. An overview of the imaging parameters used for 1.5 T imaging is provided in Table 2, while the imaging parameters used for 3 T imaging are listed in Table 3. Note that the pulse sequences for gradient-echo (GE), MARS and slice encoding for metal artefact correction (SEMAC) sequences have been adapted from the values provided in the MRI guidelines to reduce scanning time. Other spin-echo (SE), diffusion-weighted imaging (DWI) or turbo spin-echo sequences (TSE) were used in their standard settings as used in the hospital.

*Table 2: Imaging parameters for 1.5T MRI*

| Parameter | TSE | GE | MARS | SEMAC | SE |
|---|---|---|---|---|---|
| Weighting | T2 | T2 | T2 | T2 | T1 |
| Echo time (ms) | 80 | 15 | 80 | 80 | 15 |
| Repetition time (ms) | 3000 | 100 | 3344 | 3346 | 450 |
| Slice thickness (mm) | 3.5 | 4 | 3 | 3.5 | 4 |
| $B_{1+RMS}$ ($\mu$T) | 2.6 | 0.36 | 2.65 | 3.28 | 2.59 |
| dB/dt (T/s) | 32.05 | 4.09 | 91.05 | 91.94 | 7.15 |
| Flip angle (°) | 90 | 30 | 90 | 90 | 69 |
| Pixel bandwidth (Hz/px) | 395 | 109 | 465 | 438 | 109 |
| Acquisition matrix | 268x240 | 256x256 | 480x421 | 344x320 | 308x232 |
| Duration (mm:ss) | 02:42 | 01:58 | 05:01 | 07:02 | 03:24 |
| Duration compared to guidelines (mm:ss) | N/A | -07:06 | -06:48 | -05:58 | N/A |

*Table 3: Imaging parameters for 3T MRI*

| Parameter | TSE | DWI | GE | SEMAC | MARS | TSE |
|---|---|---|---|---|---|---|
| Weighting | T1 | Diffusion | T1 | T2 | T2 | T2 |
| Echo time (ms) | 11 | 83 | 9 | 80 | 80 | 80 |
| Repetition time (ms) | 531 | 4201 | 100 | 3553 | 3000 | 3000 |
| Slice thickness (mm) | 4 | 3 | 4 | 3.5 | 3 | 3 |
| $B_{1+RMS}$ ($\mu$T) | 1.99 | 1.3 | 1.87 | 1.9 | 1.96 | 1.96 |
| dB/dt (T/s) | 48.16 | 102.08 | 39.36 | 99.31 | 99.17 | 89.56 |
| Flip angle (°) | 90 | 90 | 80 | 90 | 90 | 90 |
| Pixel bandwidth (Hz/px) | 581 | 1995 | 217 | 979 | 979 | 435 |
| Acquisition matrix | 256x256 | 152x122 | 256x256 | 304x273 | 304x264 | 304x264 |
| Duration (mm:ss) | 02:33 | 00:50 | 02:10 | 05:55 | 04:30 | 02:18 |
| Duration compared to guidelines (mm:ss) | N/A | N/A | -00:52 | -00:54 | -00:19 | N/A |

All images were acquired with a body excitation coil, so scanner settings were adapted for whole-body averaged SAR. For imaging the volunteer head coils were used (15 channels in 1.5T and 32 channel in 3 T), while a combination of a torso coil and spine coil were used for imaging the phantom (32 channel in 1.5T and 32 channel in 3 T).

## Image analysis

Images were post-processed in Slicer version 4.11.2 [34] where artefact sizes were measured according to the method presented by Sharon et al. [35]: the size of each artefact was measured in the slice showing the largest artefact dimension as also prescribed in ASTM F2119-07 (2013)[36]. The artefact was then quantified by measuring. its dimensions in two mutually perpendicular directions that are parallel to the main anatomical axes. As all scans were acquired with an axial slice selection, the length of the artefact was defined in the anteroposterior direction (AP) while the width of the artefact was defined in the mediolateral direction (ML). In order to reduce variability a total of 3 consecutive measurements were made, and the average was calculated. A paired t-test was then performed to investigate the similarity in reported artefact sizes between the phantom and the volunteer. This was done using the average measurement values pooled over all test conditions for both the length and width of the artefact. In addition, a linear fit was performed between the dimensions measured using the volunteer and the phantom to quantify any consistent differences between the two.

# Results

Images acquired using the phantom show the implant-induced artefacts in an anatomically relevant setting. Figure 3 depicts images acquired using the phantom. Figura3a depicts a T2 weighted image acquired in 3T when the non-magnetic plug was in place. The image shows an implant-induced artefact around the area of the cerebellum. In addition, the image also shows the eyes and skull of the phantom. The top part of the cerebellum is appearing in black as the brain was partly filled with air. The brain was filled up with the 0.025 mM $MnCl_2$ solution before continuing the experiment using the MRI kit in the 1.5T environment. The eyes also appear to be partly filled with air, yet these were not filled during the experiment. Figure 3b depicts a T1 weighted image acquired in 1.5T with the magnet and splint in place. It depicts a large implant-induced artefact around the cortex of the model, which shows regions of white and grey matter together with the ventricles in the centre of the image.

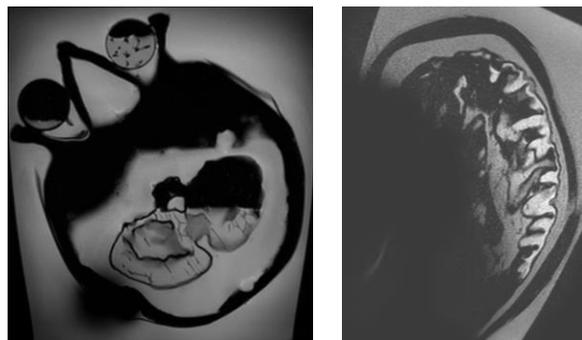

*Figure 3: Images acquired with the head phantom. (a) image showing both eyes, part of the cerebellum and the skull. (b) superior image showing the cortex, including grey and white matter regions and the ventricles*

The measured artefact sizes for all test cases are reported in Table 4 and Table 5 for 1.5T and 3T, respectively. Both tables report the mean value measured over the three consecutive measurements. The standard deviation across these measurements was on average 1.5 mm. The maximum artefact for the phantom and the volunteer appear at approximately the same anatomical location. Measurements of length differed significantly between <u>artefacts</u> measured in the phantom and the volunteer (p = 0.0011). Measurements of width did not differ significantly (p = 0.1023), yet do show a similar trend. A linear fit between artefact dimensions measured in the phantom as a function of artefact dimensions measured in the volunteer revealed that the size was consistently overestimated with about 20%, both in the length and width.

Replacing the permanent magnet with a non-magnetic plug largely reduces the dimensions of the largest artefact in the 1.5T environment. Using TSE, MARS or SEMAC sequences allows for reducing the artefact to a diameter below 5 cm when the implant magnet is removed. With the magnet and magnetic splint in place, the artefact diameter increases to 10-16 cm for these sequences. DWI and GE imaging induced larger artefacts of up to 10 cm in diameter with the magnet removed and up to 20 cm in diameter with the magnet and MRI kit in place. Artefacts induced using a regular SE sequence reside between both these optimal/worst-case dimensions. Artefacts in 3T are in general larger than those in 1.5T, yet there is no uniform scaling factor for the measured dimensions between both field strengths.

Not only the maximum artefact size is reduced, but also its location is changed. Images with a non-magnetic plug show a maximum artefact around the implantation site of the actuator, which is located more basally around the cerebellum. Images acquired with the magnet in place show maximum artefacts more superiorly around the magnet itself, affecting the visibility around the cortical regions. This is visually depicted for the volunteer test case in Figure 4. Imaging with the permanent magnet removed allows visualizing almost the entire ipsi- and contralateral brain, while imaging with the magnet in place is mostly useful for imaging when visualizing the contralateral brain.

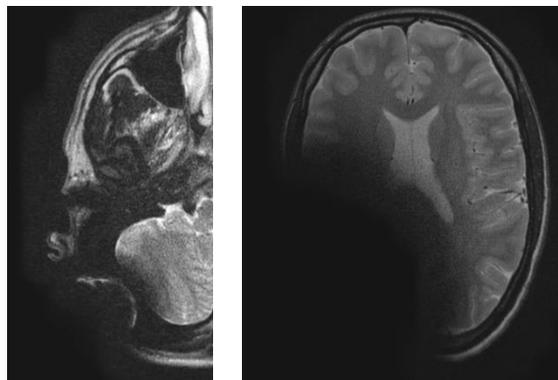

*Figure 4: Slices with maximum artefact for 1.5T MRI, illustrating the difference in artefact size and location for the non-magnetic plug (a) and the magnetic splint (b)*

*Table 4: Image artefacts measured for 1.5T imaging.*

|  |  | T2w_TSE | | T2w GE | | T2w TSE MARS | | T2w TSE SEMAC | | T1w SE | |
|---|---|---|---|---|---|---|---|---|---|---|---|
|  |  | AP (cm) | ML (cm) | AP (cm) | ML (cm) | AP (cm) | ML (cm) | AP (cm) | ML (cm) | AP (cm) | ML (cm) |
| Non-magnetic plug | Phantom | 3.9 | 3.8 | 10.3 | 9.3 | 4.8 | 4.5 | 4.7 | 4.3 | 8.0 | 6.0 |
|  | Volunteer | 4.6 | 3.6 | 8.2 | 8.5 | 3.3 | 3.7 | 2.9 | 2.4 | 6.7 | 3.8 |
| MRI kit | Phantom | 16.5 | 11.4 | N/A | N/A | 15.5 | 11.6 | 14.9 | 11.3 | 18 | 8.6 |
|  | Volunteer | 13.4 | 10.9 | 19.4 | 16.3 | 12.8 | 11.0 | 12.5 | 11.1 | 13.4 | 10.5 |

*Table 5: Image artefacts measured for 3T imaging.*

|  | T1w TSE2 | | DWI | | GE | | T2w TSE SEMAC | | T2w TSE MARS | | T2w TSE | |
|---|---|---|---|---|---|---|---|---|---|---|---|---|
|  | AP (cm) | ML (cm) | AP (cm) | ML (cm) | AP (cm) | ML (cm) | AP (cm) | ML (cm) | AP (cm) | ML (cm) | AP (cm) | ML (cm) |
| Phantom | 5.8 | 4.5 | 8.7 | 5.6 | 8.3 | 7.4 | 5.7 | 5.1 | 5.4 | 5.3 | 5.1 | 5.4 |
| Volunteer | 4.8 | 4.3 | 9.7 | 7.3 | 7.2 | 5.4 | 5.2 | 5.1 | 3.9 | 4.3 | 4.7 | 4.5 |

The ipsi- and contralateral ear canals and cochleae can be visualized in both 1.5T and 3T when the implant magnet is replaced by a non-magnetic plug. This is illustrated on Figure 5a and Figure 5c for a T2-weighted 3T MRI scan. When the magnet is in place and the MRI kit is used during 1.5T the contralateral cochlea can easily be visualized as depicted in Figure 5b and Figure 5d. Using a MARS sequence, even the ipsilateral cochlea can be distinguished with distortions as illustrated in the same figure.

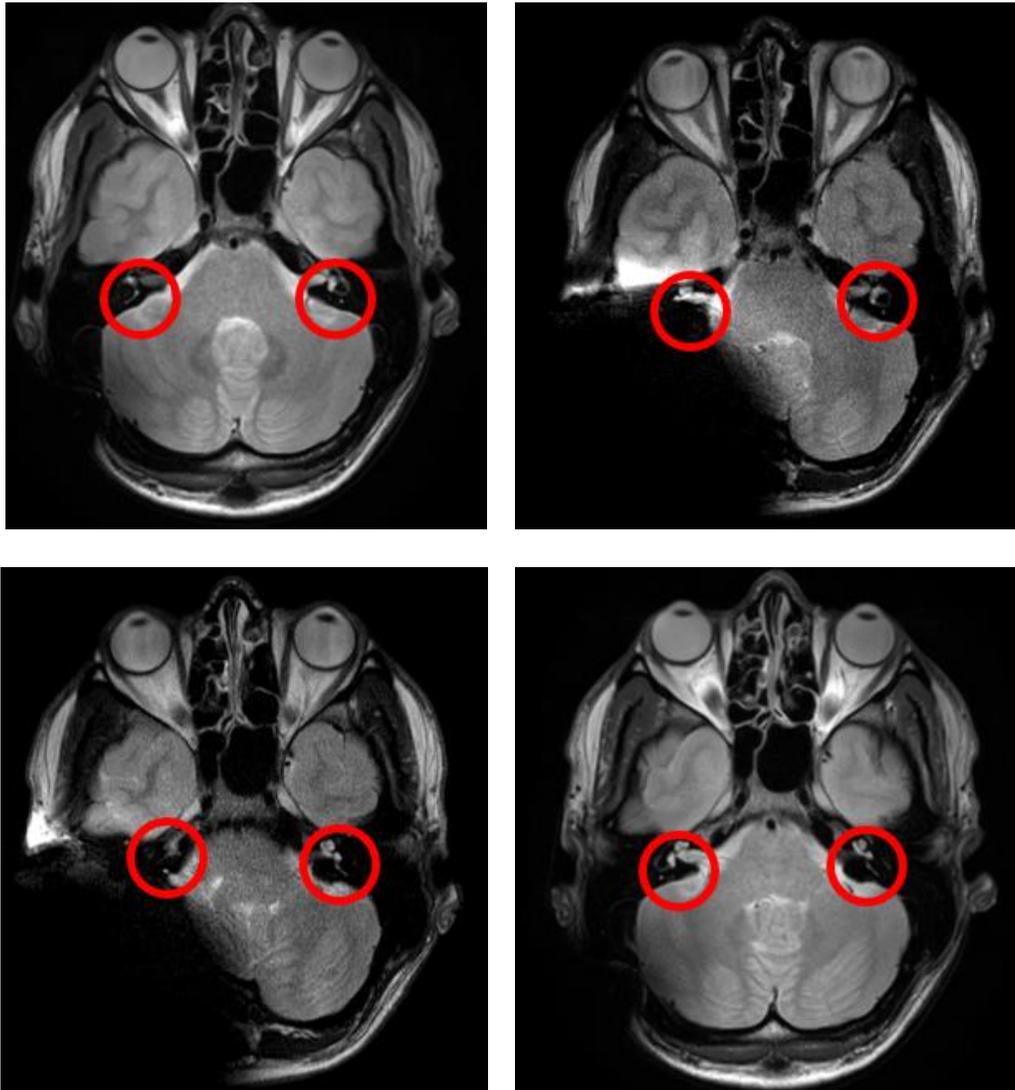

*Figure 5: The ability to show both auditory pathways. (a) a T2 weighted image acquired in a volunteer in the 3T environment, (b) depicts a similar dataset acquired during 1.5T when the magnet was left in place and the MRI kit was used. Images (c) and (d) show an inferior slice for the 3T and 1.5T images, respectively. Cochleae are encircled in red.*

## Discussion

Several anatomical phantoms have been developed in the past to investigate different aspects of MRI imaging, mainly related to either MRI-induced heating or image artefacts. The approach presented in this study focused on developing a phantom to investigate implant-induced image artefacts. In addition to bone structures and the eyes, a brain structure was developed that contains grey matter, white matter and ventricular structures. Compared to other phantoms that have been developed, the presented phantom can distinguish between the three most prominent anatomical structures (grey and white matter, ventricles) whilst other anatomical head phantoms remain mostly limited to bone structures [25] or bone structures and grey and white matter [26]. Artefacts were quantified according to the method of Sharon et al. [35], where the artefact diameters are measured in two perpendicular directions in the slice with the largest artefact. Dimensions reported in the ATBCI MRI guidelines divert from the reported values in this study. The implant manufacturer uses the methods described in ASTM F2119-07 (2013) [1], which requires measuring the artefact from the centre of the implant outwards. Artefact sizes reported in this work are therefore larger as compared to those reported in the guidelines, as the presented method reports on diameters, while the standard requires measuring radii. Despite the different measurement methods, it is clear that the artefacts are of similar size between the guidelines and this study. The presented work builds on the information provided by the manufacturer by reconfiguring the pulse sequences to limit the scan time without compromising image quality. Pulse sequences in the 3T environment that were adapted from the guidelines are up to 54 seconds shorter compared to the scan durations provided in the device guidelines. Differences in the sequences in the 1.5T scanner differ up to 7 minutes depending on the sequence.

Several studies have been published in the past where artefacts induced by the BoneBridge active transcutaneous BCI were studied. As this device is labelled to be MR conditional in the 1.5T environment, all studies have been limited to this field strength. Early work by Steinmetz et al. [17] showed in a case report that imaging the BoneBridge BCI 601 with a T2 weighted BLADE sequence does not allow evaluation of any part of the brain. Yang et al. [18] showed contradicting results where regular T1 and T2 imaging sequences were used to accurately visualize the contralateral brain. The ipsilateral internal auditory canal could also be distinguished, yet with geometric distortions. Wimmer et al. [19] implanted the BCI 601 in two cadaver heads to investigate the use of metal artefact reduction sequences. Without these sequences, they report artefact radii up to 10 cm from the implant centre, impacting the visibility of the entire ipsilateral brain and part of the contralateral brain. Artefact dimensions have not been reported for the test case where artefact reduction sequences were applied, but from the assessability study the authors have performed it can be inferred that artefact size was reduced to allow visualizing the ipsilateral cortex. The ipsilateral auditory canal, cerebellum and petrous bone remained affected by the artefact [19]. After the release of the BoneBridge BCI602, Utrilla et al. [20] performed a comparative study between both device versions. They report maximum artefact radii using a T2 weighted TSE sequence between 6.75 and 8.39 cm measured from the geometric centre of the artefact. Talon et al. [11] finally reported maximum artefact radii of 6.5 to 7.0 cm depending on the implant position in the anatomy using a SEMAC-VAT WARP sequence in cadaver heads. When comparing these results with the artefact dimensions presented in this work, it is clear that artefact sizes in 1.5T are importantly reduced. With the permanent magnet and MRI kit in place, a SEMAC sequence creates artefact radii (i.e. half of the values reported) of up to 7.5 cm, whilst using a non-magnetic plug leads to an artefact radius of up to 2.3 cm. Using the T2 weighted TSE sequence, artefact radii of up to 8.2 cm are measured with the magnet and MRI kit in place, whilst artefact radii of up to 2.3 cm are measured with the non-magnetic plug in place. An important note is also that the location of maximum artefact differs between the devices. With the non-magnetic plug, the maximum artefact dimension is around the implantation site of the actuator, whilst the maximum artefact dimension with the magnet in place is located superiorly for the Osia device. This implies that the artefact can be reduced importantly by removing the magnet in case the entire ipsilateral brain needs to be visualized.

The physical principle behind the manifestation of artefacts lies in the fact that the main $B_0$ magnetic field is disturbed locally due to local differences in magnetic susceptibility [37]–[39]. Field inhomogeneities are mostly expressed in parts per million (ppm), with biological variations typically up to 3.2 ppm in tissue and 9 ppm at air/tissue interfaces [38]. The differences are orders of magnitude larger around metallic implants, with reported values up to $10^{11}$ ppm [38], [39]. In the present study, the largest artefact sizes were found when the magnetic MRI kit was used to keep the implant magnet in place. A bench measurement of the induced magnetic field of the magnetic MRI kit combined with the implant magnet showed a field strength of 3.3 mT or 2200 ppm at a distance of 3 cm. Considering a roughly $1/r^2$ distance dependence of this induced magnetic field, this image disturbing field would be reduced to 50 ppm at 20 cm distance. It should be noted that susceptiblity artefacts can be reduced to a certain extent by increasing the pixel bandwidth [38] and shimming. As mentioned above, the image disturbance measured in this work extended over about 10 cm.

With respect to imaging of the auditory pathway, the presented results show that both cochleae can be visualized in both 1.5 T and 3 T, and 1.5 T with the magnet in place. In the latter case, the ipsilateral auditory canal can not easily be distinguished.

As is the case in every study reporting implant-induced artefacts, it is mainly the approximate artefact size and location that are of importance and not the absolute values. The exact dimensions of an artefact are impacted by the complete scanner settings, the coils used, the artefact measurement method and the position/orientation of the test subject. As this study has shown, the absolute dimensions of the artefact differ significantly between the phantom and the volunteer. In part, this can be attributed to the difference in orientation between both specimen. The volunteer was placed inside the scanner supine, while the phantom was placed into the scanner at a 45° degree axis relative to the supine position. The difference in head orientation creates an identical difference in implant orientation, which not only potentially results in differences in magnetization between both cases but also creates a difference in length/width orientation with respect to the $B_0$-field. Nonetheless, both the phantom and the volunteer provide relevant estimates of both the artefact location and size.

Acquiring images to study implant-induced artefacts has been performed by other authors, providing important insights into the location of extent of image artefacts[14], [33]. Future studies could investigate what the impact is of having an implant placed on the head versus under the skin. This can be done by performing a study within the same subject before and after implantation of a device.

The presented study mainly comprised of a proof-of-concept of an anthropomorphic phantom, consisting of a comparative case study with a volunteer. Images acquired are therefore mainly focused on clinical guidelines and clinical value. A limitation of the study is that the description of the phantom's anatomical accuracy is limited by the image resolution. Increasing the resolution of images would require exceeding the manufacturer's recommended guidelines, which was out of scope for this study. A future study could comprise of acquiring images with maximum resolution, to distinguish and identify all relevant present micro-anatomical structures in the brain. The current study however already indicates the potential of the phantom as a preclinical tool for image artefact studies. In order to allow using the same coil hardware between a volunteer and the phantom, a cylindrical housing for the phantom can be provided in future studies. The impact of scanning with different hardware is however expected to be limited, as $B_1$ and $B_0$ shimming ensure a maximally homogeneous field for imaging a region of interest.

In summary, an anthropomorphic head phantom has been presented that was designed to allow the creation of realistic images of implant-induced artefacts at a pre-clinical stage. The phantom allows for assessing the most important macroscopic brain structures, yet more research is necessary to look into the potential of the phantom to also allow visualizing microscopic brain structures. The phantom was used in a comparative study with a volunteer using a piezoelectric active transcutaneous bone conduction implant. The study was performed in both a 1.5T and 3T environments, and results show that when the scanner settings are properly defined, visualization of almost the entire brain (i.e. including both hearing pathways) is feasible.

# Supplementary materials to "Metal artefact reduction sequences for a piezoelectric bone conduction implant using a realistic head phantom in MRI" by Fierens et al

## Supplementary document 1- Relaxivity curves for $MnCl_2$ and $CuSO_4$

*Table 6: Concentration (g/l) for containers filled with a $CuSO_4$ solution.*

| Identifier | Concentration (g/l) |
|------------|---------------------|
| 1 | 2 |
| 2 | 1 |
| 3 | 0.5 |
| 4 | 0.25 |
| 5 | 0.125 |
| 6 | 0.0625 |

*Table 7: Concentration (mM) for containers filled with a $MnCl_2$ solution.*

| Identifier | Concentration (mM) |
|------------|--------------------|
| 1 | 0.2 |
| 2 | 0.1 |
| 3 | 0.05 |
| 4 | 0.025 |
| 5 | 0.0125 |
| 6 | 0.00625 |

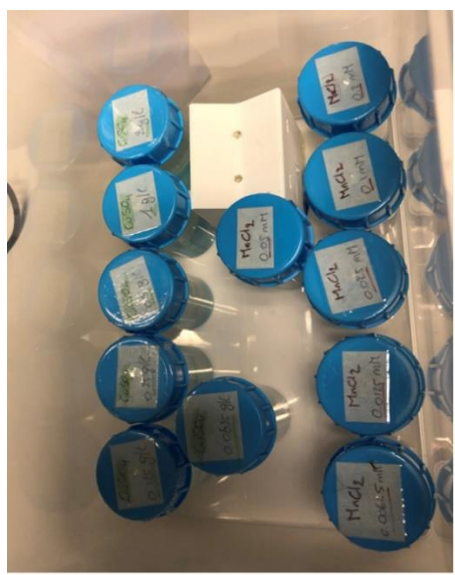

**(a)**

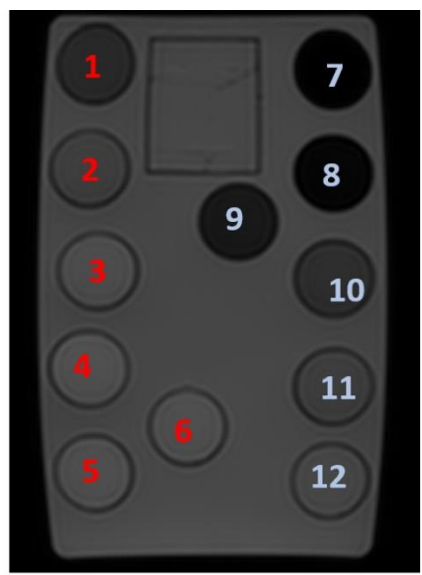

**(b)**

*Figure 6: container placed in a water bath (a) and their identifiers*

## Relaxivity model

The 2-parameter model to calculate the T1 relaxation time per solution was defined by Thangavel and Saritas[1] as:

$$S = S_0(1 - 2e^{\frac{-TI}{T1}} + e^{\frac{-TR}{T1}})$$

with $S$ the signal intensity in pixels of interest [$a.u.$], $S_0$ the background pixel intensity, TI the inversion time ($ms$), and TR the repetition time ($ms$). Once the T1 values are defined for each concentration $C$ [$mM$], the relaxation rate $R1$ [$1/s$] can be calculated as:

$$R_1(C) = \frac{1}{T1}(C) = r_1 C + C_0$$

with $r_1$ the relaxativity [$mM\text{-}1s\text{-}1$] and $C_0$ the background concentration [$mM$].

## T1 relaxivity curves

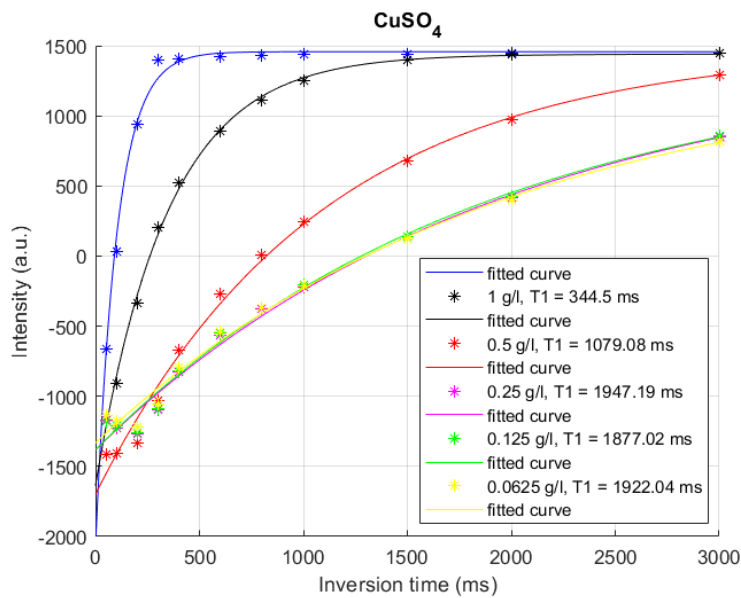

*Figure 7: Calculated T1 times for CuSO4 (ms).*

---

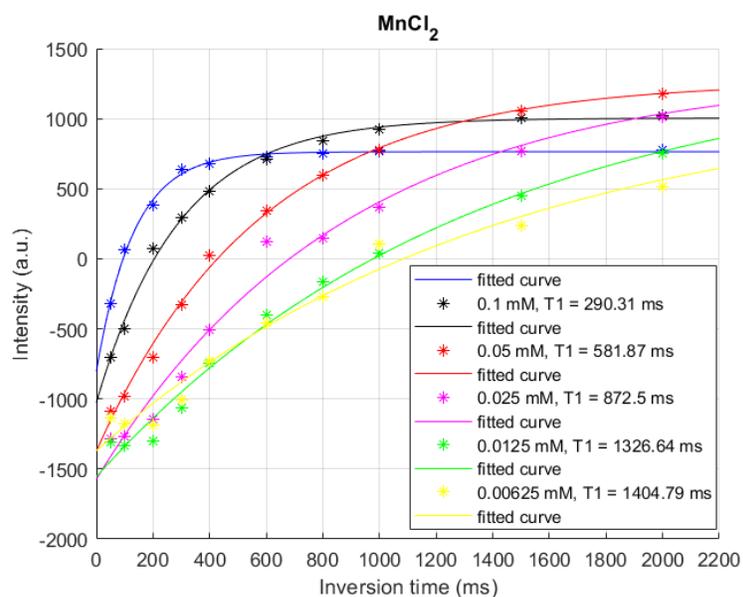

*Figure 8: Calculated T1 times for MnCl2 (ms).*

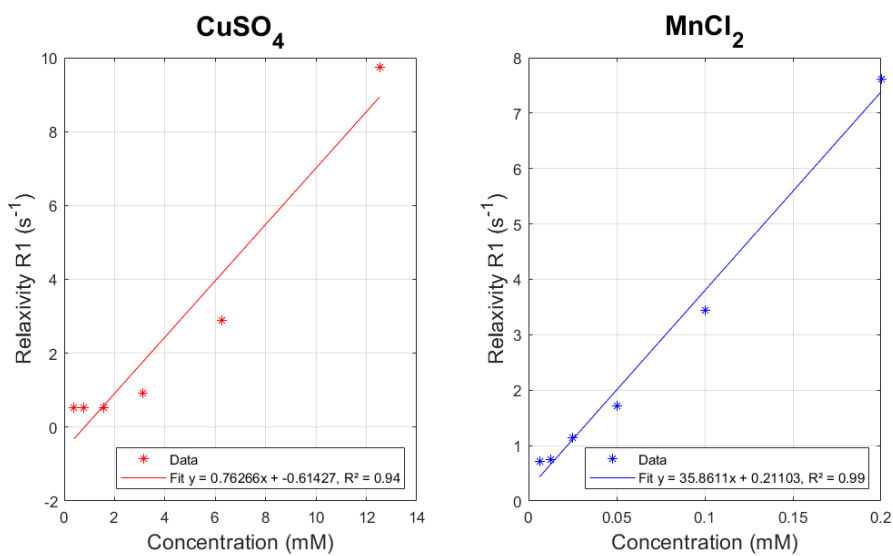

*Figure 9: R1 relaxivity for MnCl2 and CuSO4 solutions.*

*T2 relaxivity curves*

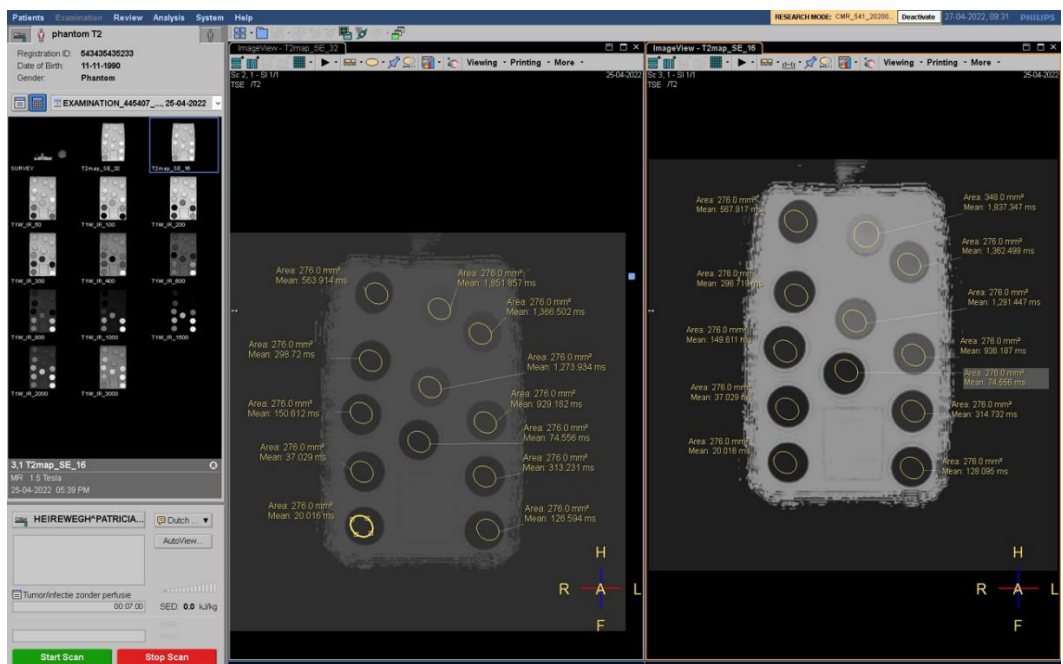

*Figure 10: T2 relaxation times*

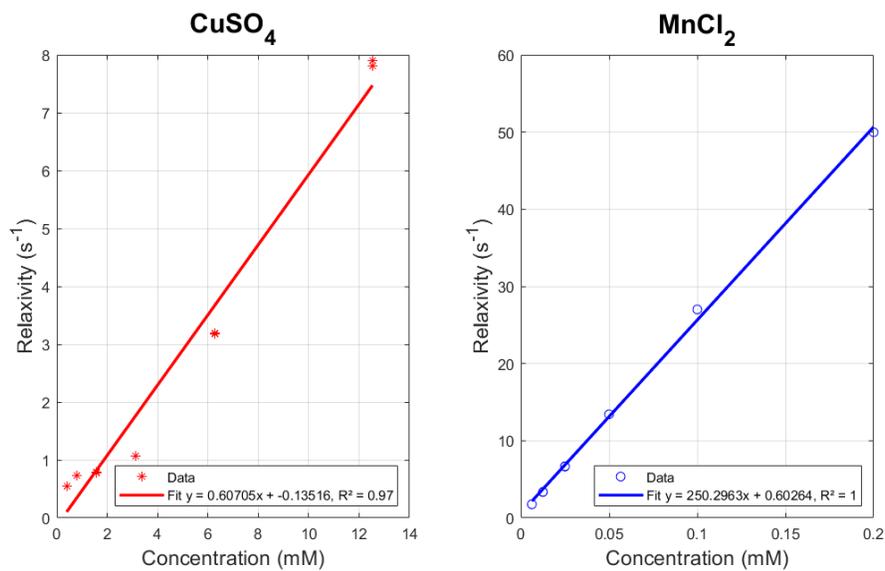

*Figure 11: R2 relaxivity for MnCl2 and CuSO4 solutions.*

# Supplementary document 2- Relationship between artefact sizes measured in the volunteer versus the phantom

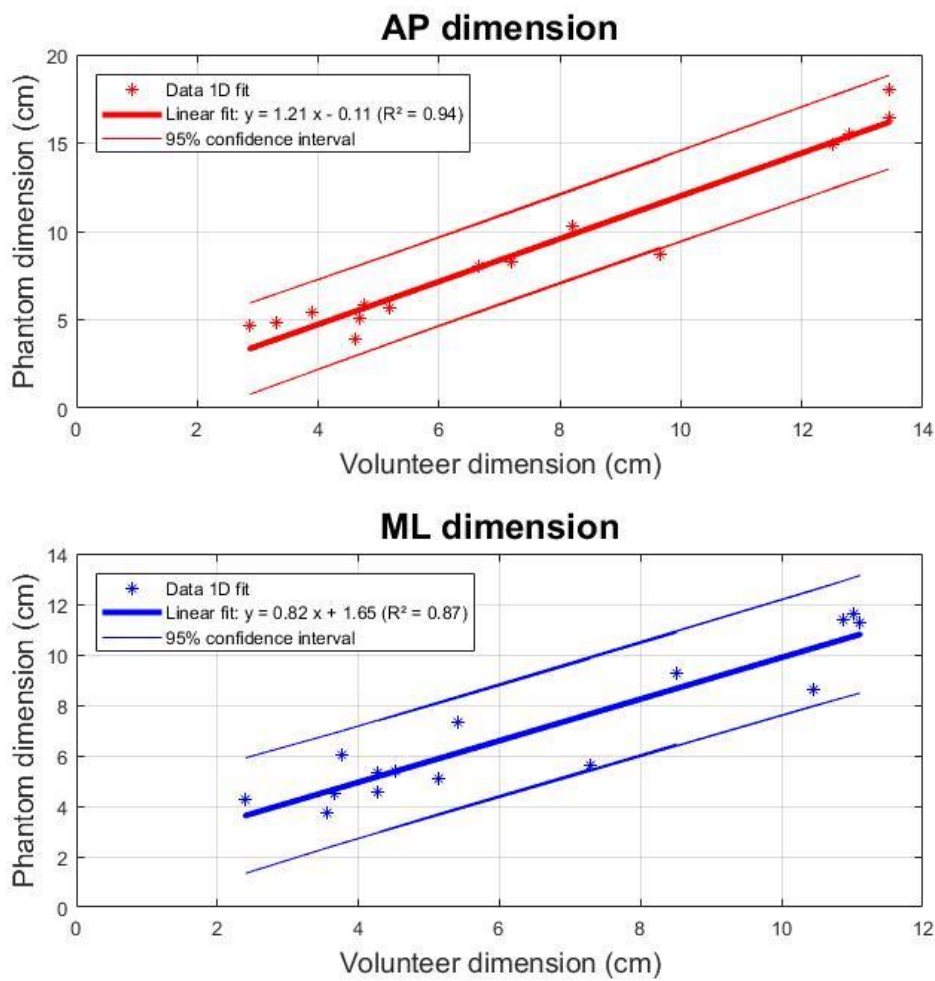

*Figure 12: Linear fit for artefact length (top) and width (bottom) where artefact sizes for the phantom are defined as a function of those measured in the volunteer in images acquired using the same pulse sequence.*